\documentclass[twocolumn,8pt]{extarticle}
\usepackage[utf8]{inputenc}
\usepackage{wrapfig}
\usepackage{graphicx}
\usepackage{algorithmic}
\usepackage{algorithm}
\usepackage{subcaption}
\usepackage{authblk}

\author[1]{Pedro Moreno}
\author[2]{Ricardo Rocha}
\affil[1,2]{CRACS \& INESC TEC, Department of Computer Science, Faculty of Sciences, \authorcr
University of Porto \authorcr
Rua do Campo Alegre s/n, 4169-007 Porto, Portugal}
\affil[1]{pmoreno@dcc.fc.up.pt}
\affil[2]{ricroc@dcc.fc.up.pt}
\title{\huge\textbf{Releasing Memory with Optimistic Access: A Hybrid Approach to Memory Reclamation and Allocation in Lock-Free Programs}}
\date{}

\begin{document}
\maketitle

\begin{abstract}

Lock-free data structures are an important tool for the development of
concurrent programs as they provide scalability, low latency and avoid
deadlocks, livelocks and priority inversion. However, they require
some sort of additional support to guarantee memory reclamation. The
Optimistic Access (OA) method has most of the desired properties for
memory reclamation, but since it allows memory to be
accessed after being reclaimed, it is incompatible with the traditional
memory management model. This renders it unable to release memory to
the memory allocator/operating system, and, as such, it requires a
complex memory recycling mechanism. In this paper, we extend the
lock-free general purpose memory allocator LRMalloc to support the OA
method. By doing so, we are able to simplify the memory reclamation
method implementation and also allow memory to be reused by other
parts of the same process. We further exploit the virtual memory
system provided by the operating system and hardware in order to make
it possible to release reclaimed memory to the operating system.

\end{abstract}

\section{Introduction}

With the recent developments in computer hardware focusing on the
increase of parallelism as the main way to improve performance, it is
key to have accompanying software capable of taking advantage of such
hardware. Lock-free data structures provide one of the most
fundamental building blocks for concurrent/parallel software, as the
lock-freedom property promotes scalability and guarantees immunity to
livelocks, deadlocks and priority inversion~\cite{herlihy11}. However,
in comparison to their lock-based counterparts, lock-free data
structures require additional support in order to manage memory
reclamation. This can be delegated to a garbage collector, if the
programming runtime being used provides one, but such a garbage
collector is usually not lock-free causing the system as a whole to
lose the lock-freedom property~\cite{petrank12}.

An alternative is to use a specific memory reclamation method. The
most common methods, such as \emph{pass the buck}~\cite{herlihy05} and
\emph{hazard pointers}~\cite{michael04}, are based on the idea of
threads advertising their coordinates in order to prevent other
threads from reclaiming the memory they are using. This idea however
requires every thread to constantly write its coordinates to memory
and perform expensive memory barriers in order to ensure that such
memory writes are visible. More sophisticated methods try to amortize
the memory writes and consequent memory barrier usage. Some examples
are \emph{drop the anchor}~\cite{braginsky13}, \emph{hazard
eras}~\cite{ramalhete17}, \emph{interval based
reclamation}~\cite{wen18}, and \emph{hazard hash and
level}~\cite{moreno21}, among others. Dice~et~al.~\cite{dice16} also
provide a mechanism to reduce the cost of memory barriers, but such
mechanism requires hardware/operating system support.

Instead of having threads advertising their coordinates, a more recent
strategy, called \emph{optimistic access} (OA)~\cite{cohen15oa},
allows threads to optimistically access the memory they are traversing
and only after verify if the access is valid. In order to be able to
check the validity of a memory access, the OA method moves the
responsibility to the reclaiming threads to advertise that memory
reclamation has occurred. This no longer requires threads to
constantly write to memory to advertise their locations, but only to
do extra reads to check if memory reclamation has occurred. These
extra reads are inexpensive, as they will target a cached memory
location most of the time, and require less expensive memory barriers.

An important disadvantage of existing OA based methods is that they
are unable to release memory to the memory allocator/operating system.
This happens due to the fact that, at anytime, a thread may read
memory that has already been reclaimed. To work around this problem,
these methods implement a recycling mechanism to manage the memory
being used. However, this prevents the memory used in this manner from
being reused in other parts of the same process and from being
released to the operating system.

In this work, we propose a solution to this problem without having to
make the whole application aware of the memory reclamation method. Our
proposal is to extend \emph{LRMalloc}~\cite{leite19}, a lock-free
general purpose memory allocator, in such a way that we can guarantee
memory allocations to be readable even after we free such allocations.

No guarantees are given about the content of the memory, or how it is
reused by the rest of the application. This is a good match for OA
because it already ensures that the contents of reads on possibly
reclaimed memory are to be ignored, and that memory to be written is
protected from reclamation by the use of hazard pointers.

We start by solving the problem at the memory allocator level, by
adapting LRMalloc such that it does not release memory used by the OA
method back to the operating system. This allows us to simplify the
implementation of the OA memory reclamation method as we no longer need a
recycling mechanism in order to manage the distribution of memory
between threads. This task is now covered by the memory allocator as
it was designed for this task in a general sense. We also gain the
ability to reuse memory reclaimed by the OA method across the whole
process. As we will see, all this is possible with minimal changes to
the LRMalloc memory allocator.

Then, to complete our solution, and have the ability to release the
memory used by the OA method to the operating system, we exploit how
current operating systems/hardware use virtual memory. As we need the
virtual addresses (pages) to remain accessible after they have been
used by the OA method, but we do not care about the contents on the
physical memory (frames) they are mapped to, we map all these multiple
pages to the same frame. This allows us to free all the frames our
pages were previously mapped to while keeping the pages still valid
for access.

Modern operating systems apply similar strategies, e.g., when a memory
request is made to the operating system, no frame is immediately
reserved, only the pages are made valid by being all pointed to a
single \emph{copy on write} zero filled frame. Only when a memory
write is attempted in these pages, is that the operating system copies
the zero filled frame to a new free frame and maps the page to it.
This all happens transparently to the application, which never
notices that the memory given to it at the start was not
actually backed by physical memory. One of the strategies we propose
to implement the remapping of pages exploits this operation system
behavior, while the other strategy will do the remapping in a more
explicit fashion using the shared memory mechanisms of current
operation systems.

The remainder of the paper is organized as follows. First, we
introduce relevant background. Then, we present in detail the main
ideas supporting our approach and discuss its current limitations.
Next, we show a set of experiments comparing our model against the
original OA method. At the end, we present conclusions and further
work directions.


\section{Background}

This section briefly introduces relevant background about virtual
memory and memory allocation systems and describes in more detail the
LRMalloc memory allocator and the Optimistic Access (OA) method.


\subsection{Virtual Memory}

Virtual memory is a memory management system that works as an
abstraction layer that allows for a multitude of optimizations in
modern operating systems. The main idea is to have a translation layer
between the memory addresses viewed by a user process and the actual
physical addresses in main memory. The translation is done in hardware
by the \emph{memory management unit} (MMU) and relies on a cache named
\emph{translation lookaside buffer} (TLB). This introduces an
overhead, as with virtual memory, when trying to access a memory
location, one first needs to consult where the virtual address resides
in physical memory. This requires extra memory accesses in order to
obtain the physical memory location, however by the use of an
efficient TLB this disadvantage is mostly mitigated. Modern systems
define the granularity of a page/frame to be a power of 2, usually between
4KiB and 1GiB total size.

The main benefits provided by virtual memory are the ability for
processes to oversubscribe memory allowing them to use more memory
than what is physically available, the ability of multiple processes
having the same address space, the ability to move unused pages from
memory to persistent storage when under memory pressure, and the
ability to block a process from accessing or modifying any memory that
does not belong to it. Virtual memory also allows memory to be shared
between processes, the most common case being shared libraries, so
multiple processes can use the same copy of a library in physical
memory but each have it in a different memory address. Another
important use case is efficient inter-process communication, made
possible having two or more processes mapping a single region of
physical memory into their own address spaces.

Further optimizations include the ability to only load frames when
they are needed, meaning that when a process is loaded into memory, it
does not need to be entirely loaded, only the necessary frames are
loaded as the corresponding pages are accessed. For example, an error
routine that is never called would never actually be loaded into
physical memory. When a process requests memory from the operating
system, a similar optimization can be done, every page the process
requests can be initially mapped to a single zero filled frame and
only mapped to free memory frames when they are actually written to.
As we will see later, this is one of the features that we will take
advantage of for our proposal.


\subsection{Memory Allocation}

Memory allocators serve as an interface between processes and the
operating system, satisfying memory requests of any size in such a way
that processes waste as little additional memory and time as possible.
To do so, a memory allocator starts by acquiring pages from the
operating system that are then subsequently divided to satisfy smaller
allocation requests, and later combined in order to give complete
pages back to the operating system. Classic memory
allocators~\cite{wilson95} tended to use strategies like
\emph{best-fit}, in which they find the smallest block of contiguous
memory that can satisfy the request and, if such a block is still too
big, it is split to the right size so they can keep what remains to a
future allocation. Another strategy is \emph{first-fit}, in
which instead of finding the smaller continuous block that satisfies a
request, they simple use the first block found. This strategy has a
speed advantage, but can increase memory waste.

A more modern strategy is to use \emph{size classes}, where any
request is met by rounding up to the nearest size class. Blocks of a
size class are created by splitting a bigger block into many blocks of
the same size. The size classes need to be carefully selected,
therefore avoiding too many different classes and possibly allocations
of large blocks that result in a limited amount of allocations from
it, or too few classes and possibly wasting memory by having to
provide a much larger allocation than needed due to the nonexistence
of a large enough smaller size. Size classes are very time efficient
and tend to improve memory locality, therefore also improve the global
performance of applications beyond memory allocation.

With the advent of multi-core processors, in order to further improve
performance and scalability, different proposals were adopted to
minimize the amount of synchronization between threads. These gave
origin to mechanisms such as \emph{private heaps}~\cite{berger00}, in
which each thread has a private allocator implementing specific
strategies to deal with frees that occur in threads different from the
one where the memory was allocated. These strategies can be used to
kept the free memory in the thread in which it was freed until it is
allocated again; to immediately give back the free memory to the
thread it was allocated on; or to give back only after a threshold is
met. An alternative mechanism is to use a per thread cache on top of a
shared heap~\cite{lee14}.


\subsection{LRMalloc}

\begin{figure}
\centering
\includegraphics[width=4.5cm]{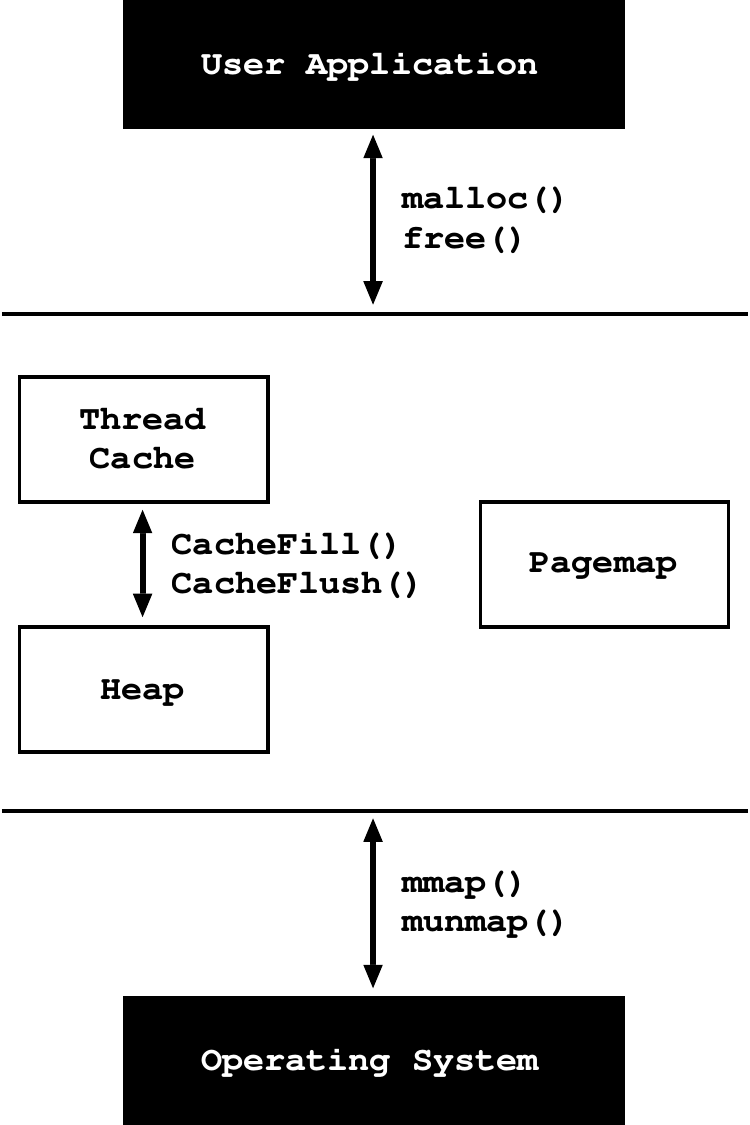}
\caption{LRMalloc's overview}
\label{fig_lrmalloc}
\end{figure}

LRMalloc~\cite{leite19} is a modern lock-free memory allocator that
uses size classes and thread caches as described above. It has three
main components: (i) the \emph{thread caches}, one per thread; (ii)
the \emph{heap}; and (iii) the \emph{pagemap}.
Figure~\ref{fig_lrmalloc} shows the relationship between these three
components, the user’s application and the operating system,

The thread caches use a stack for every size class, so that a memory
request becomes simply a pop on the corresponding size class stack,
and a memory free becomes a stack push. When a memory request is made
and the corresponding stack is empty, then the stack is filled from
the heap, and when a memory free happens and the stack is full, it is
flushed back to the heap. The size of the stack is limited in order to
prevent \emph{blowup}~\cite{berger00}. The caches are local to a
thread, so they only synchronize with other threads when a fill or
flush from/to the heap occurs.

The heap is responsible for managing \emph{superblocks}, which are
large blocks of memory obtained from the operating system that are
then divided into blocks of a size class to be given to the thread
caches. Superblocks are managed through \emph{descriptors}, an object
that contains the superblock metadata and that is never reclaimed.
When a superblock is released to the operating system, the associated
descriptor is added to a recycling pool in order to be
reused for a future superblock. The descriptor contains information,
such as, where the superblock begins, its associated size class, the
number of blocks it possesses, the index of the first free block and
the number of free blocks.

Superblocks can be in one of three states: (i) \emph{full}, if all its
blocks are in use; (ii) \emph{empty}, if all its blocks are available
for allocation; or (iii) \emph{partial}, if it has available and
allocated blocks. The initial state of a superblock is always full, as
all its blocks are immediately used to fill a cache. Then it becomes
partial as some blocks are returned to it by cache flushes, at which
point it can either become full again, if a threads uses it to fill
its cache, or it can become empty, if all blocks are returned to
it. When a superblock becomes empty, it cannot be used again and its
memory is released back to the operating system. When threads try to
fill their caches they give priority to partial superblocks and, if
none is available, a new superblock is created by requesting memory
from the operating system.

The pagemap is a simple lock-free data structure that stores metadata
for each page in use. Taking into account that superblocks are always
aligned with pages and have a size that is a multiple of the page
size, blocks in the same page always belong to the same superblock. So
this metadata includes the superblock that a page belongs to and its
associated descriptor. The main usage of the pagemap is to allow
finding the corresponding superblock for a block that is flushed from
the cache, or to allow finding the appropriate cache (with the correct
size class) when memory is receive from the application through a call
to the \emph{free()} procedure.


\subsection{Optimistic Access}

A memory reclamation method for a lock-free data structure is a
mechanism that detects when an node removed from the data structure
can no longer be referenced by any running thread, and thus uses such
information to free the corresponding memory to the memory
allocator/operating system. Usually, such methods require some sort of
validation to avoid accessing memory that has been already reclaimed.

An alternative approach is the one followed by the \emph{optimistic
access} (OA) method~\cite{cohen15oa}, which, as the name implies,
allows memory accesses before making sure the memory has not been
reclaimed, and only then checks the validity of the access by reading
a specific \emph{warning-bit}. If the access corresponds to reclaimed
memory, the result is ignored and the procedure is restarted from a
memory location known to be valid. However, modifying operations
cannot be performed in an optimistic manner as an optimistic CAS could
incorrectly succeed due to an ABA problem~\cite{dechev10}. For that,
OA uses a hazard pointer strategy, so before performing any atomic CAS
(\emph{Compare-and-Swap}) update operation, it first protects all
memory addresses involved by assigning hazard pointers to them and
then performs a single additional validity check by reading the
\emph{warning-bit}, therefore ensuring that the memory was valid when
it was protected by the hazard pointers. These hazard pointers are
then used to prevent the recycling of the memory they are assigned to.

The OA memory recycling mechanism is composed by three pools: (i) the
\emph{ready pool} that contains all the nodes ready to be allocated,
(ii) the \emph{retire pool} to which nodes are added when they are
retired from the data structure, and (iii) the \emph{processing pool}
that holds the nodes that are in the process of being recycled. The
recycling mechanism works in phases, and a new phase is triggered when
the ready pool is exhausted. At the start of a new phase, the
nodes present in the retire pool before the phase stars are moved to the processing pool.
Next, all threads are informed of the current recycling by their
\emph{warning-bit} being set. Finally, the nodes in the processing
pool that are protected by hazard pointers are moved back to the
retire pool, the ones not protected are moved to the ready pool.
Threads that try to retire an node during the process of moving nodes
from the retire pool to the processing pool need to help finish the
move before retiring the node.  Threads that try to start a new
recycling phase while one is already in progress need to help finish
the current phase before starting a new one.

While the recycling mechanism is complex and time consuming, it is
rarely executed, which mitigates its cost. For the more frequent
operations, such as the traversal of the data structure, this method
only needs to perform an extra read per node traversed instead of a
write, as it is the case for the hazard pointers memory reclamation
method, and it also requires a much less expensive memory barrier,
which in \emph{total store ordering} (TSO) architectures like x86-64,
translates to a simple compiler barrier and no additional hardware
instructions are emitted. Also note that writing operations only
require one validity check for setting multiple hazard pointers and
consequentially only one expensive memory barrier, compared to the
hazard pointers method which requires one per node. These
characteristics make the optimistic access memory reclamation method
extremely efficient and performant compared to the state-of-the-art,
while also having low memory bounds and not requiring any specific
support from the operating system.

A consequence of allowing optimistic accesses to possibly reclaimed
nodes is that nodes need to remain accessible after being reclaimed.
However, there is no need for the contents of the node to be
maintained, as the result of the access will be ignored in the case it
was invalid. To ensure the nodes are accessible after
being reclaimed, the recycling mechanism is used, which allows nodes
to be reused, but never released to the memory allocator or the
operating system.


\section{Our Approach}

In this section, we start by introducing how we make LRMalloc
compatible with the OA memory model and how we can use it to simplify
the OA method. Next, we present how we can exploit virtual memory in
order to allow memory to be released to the operating system.


\subsection{Memory Recycling at the Allocator Level}

As mentioned before, in a program using a lock-free data structure in
combination with the OA memory reclamation method, the memory
reclaimed can be reused by the data structure but it cannot be reused
by other parts of the program, at least without extensive
modifications both to the memory reclamation method and to the rest of
the program.

Our solution is to handle this restriction at the memory allocator
level by making sure that memory can be accessed even after being
freed. The allocator would still not provide any guarantees of the
contents of the freed memory, and we would not be allowed to write to
it either, as it could lead to corruption if such memory was already
reused in another allocation. To achieve this we extended LRMalloc
with a new function that we named \emph{palloc()} (\emph{persistent
alloc}).

\begin{figure}[ht]
	\begin{subfigure}{0.49\linewidth}
		\includegraphics[width=0.95\columnwidth]{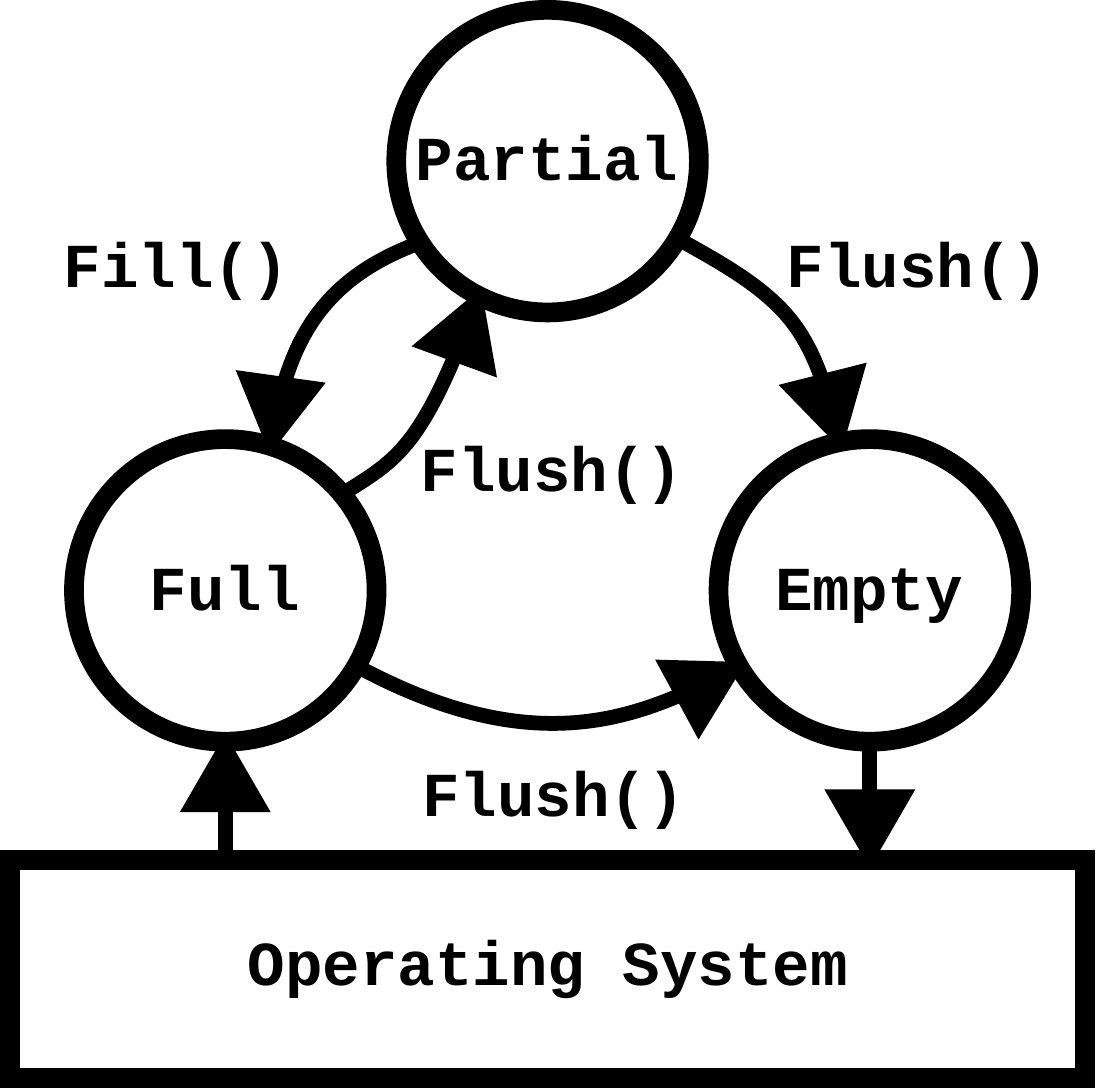}
	\caption{Non-persistent}
		\label{fig:statesa}
	\end{subfigure}
	\begin{subfigure}{0.49\linewidth}
		\includegraphics[width=0.95\columnwidth]{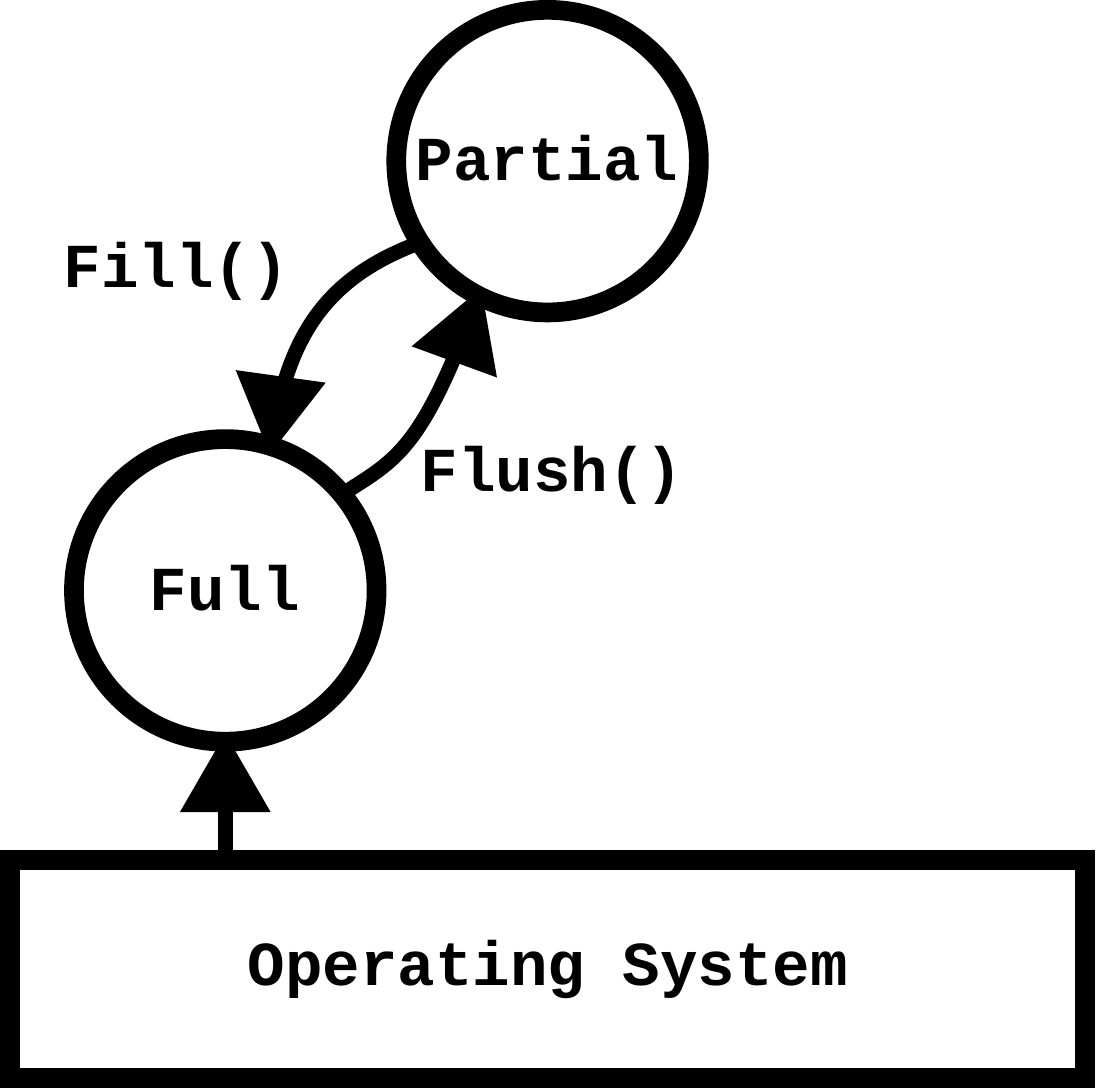}
	\caption{Persistent}
		\label{fig:statesb}
	\end{subfigure}
	\caption{State diagram for superblocks}
	\label{fig:sbstates}
\end{figure}

To implement \emph{palloc()}, we follow the same process as regular a
allocation,
but the \emph{superblock} that contains the memory block being
allocated is marked as \emph{persistent}. This mark is then used to
guarantee that persistent superblocks never reach the \emph{empty}
state, even if all its blocks are available. This change ensures that
memory allocated with \emph{palloc()} is never released to the
operating system, but can still be reused by future allocations
anywhere on the same process. Figure~\ref{fig:sbstates} shows the
state diagram for superblocks before and after being marked as
persistent.

By having an allocator that satisfies these properties, we can now
extensively simplify the memory reclamation method. As we no longer
need the memory recycling mechanism employed originally in the OA
method, we can use a much simper mechanism, similar to the one used by
the hazard pointers memory reclamation method, as shown in
Alg.~\ref{alg:wb}.

\begin{algorithm}
	\caption{Retire(Node N)}
	\label{alg:wb}
	\begin{algorithmic}
		\STATE LimboList.add(N)
		\IF{LimboList.full()}
			\FOR{T in Threads}
				\STATE T.warning\_bit.set()
			\ENDFOR
			\STATE MemoryBarrier()
			\FOR{T in Threads}
				\STATE HPSet.add(T.hazard\_pointers)
			\ENDFOR
			\FOR{M in LimboList}
				\IF{\NOT HPSet.contains(M)}
					\STATE LimboList.remove(M)
					\STATE Free(M)
				\ENDIF
			\ENDFOR
			\STATE HPSet.reset()
		\ENDIF
	\end{algorithmic}
\end{algorithm}

The idea is as follows. When a node is retired, we add it to the
reclaiming thread's \emph{limbo list}, and when the list's size
reaches a certain threshold, we perform the reclamation
procedure. During such procedure, we only need to set all the other
threads' \emph{warning-bit} and then free all nodes that are not
protected by a hazard pointer.

This mechanism however is not ideal for data structures with long
chains, such as linked lists, since as we trigger more warnings, more
restarts are needed. These restarts are inexpensive on data structures
with short chains, such as hash tables, but not so much in linked
lists, not only because the amount of work lost by a restart is high,
but also because the beginning of the chain is most likely out of the
L1 cache by the time of the restart.

To mitigate this issue, we implemented another warning mechanism that
is based on the one used in the \emph{Version Based Reclamation} (VBR)
method~\cite{sheffi21}. In this mechanism instead of having a warning
bit per thread, we have a monotonic global variable that we increment
when we want to send a warning to all threads, and threads check for
the warning by comparing the last value seen in the global variable
with the current value. With this mechanism we can allow threads to
piggy back of each other warnings, as we can forego sending a warning
if one has happened since the time the nodes we want to reclaim was
retired. Note that we not only take advantage of other threads
warnings when we see the increment in the global variable, but also
when we try to increment it with a CAS and it fails, what means that a
warning was successfully fired by another thread and we can take
advantage of it. Algorithm \ref{alg:gc} shows the retire procedure and
how it is able to piggy back of other threads. Note that this is not
possible on the previous method as the warnings are not atomic with
one \emph{warning-bit} per thread. In this algorithm, the \emph{GlobalClock}
variable represents the global monotonic variable, the \emph{LocalClock} is
a local variable used to store the last seen value of the global variable, and
the \emph{LastRetireTime} is a local variable used to take advantage of the
other threads warnings.

\begin{algorithm}
	\caption{Retire(Node N)}
	\label{alg:gc}
	\begin{algorithmic}
		\IF{LimboList.full()}
			\IF{LastRetireTime = LocalClock}
				\STATE CAS(GlobalClock, LocalClock, LocalClock + 1)
				\STATE LocalClock $\Leftarrow$ GlobalClock
			\ENDIF
		\ENDIF
		\IF{LastRetireTime $<$ LocalClock \AND LimboList.size() $>$ X}
			\STATE MemoryBarrier()
			\FOR{T in Threads}
				\STATE HPSet.add(T.hazard\_pointers)
			\ENDFOR
			\FOR{M in LimboList}
				\IF{\NOT HPSet.contains(M)}
					\STATE LimboList.remove(M)
					\STATE Free(M)
				\ENDIF
			\ENDFOR
			\STATE HPSet.reset()
		\ENDIF
		\STATE LastRetireTime $\Leftarrow$ LocalClock
		\STATE LimboList.add(N)
	\end{algorithmic}
\end{algorithm}

As mentioned earlier, with this method we end up with memory that we
can never release to the operating system throughout the lifetime of
the process. In the case that a large amount of memory is allocated
with \emph{palloc()}, that memory will continue in the process even if
the amount of memory it requires for the remainder of its lifetime is
much lower. The main advantage of this mechanism is that it requires
no additional features from the operating system or hardware compared
to any other lock-free memory allocator.


\subsection{Using Virtual Memory}
\label{sect:uvm}

Now that we have made the memory allocator compatible with the
optimistic access model, we next focus on the interaction with the
operating system. Remember that the memory allocator cannot release
superblocks marked as persistent to the operation system because they
need to remain accessible.

\begin{figure}[ht]
	\begin{subfigure}{0.49\linewidth}
		\includegraphics[width=0.95\columnwidth]{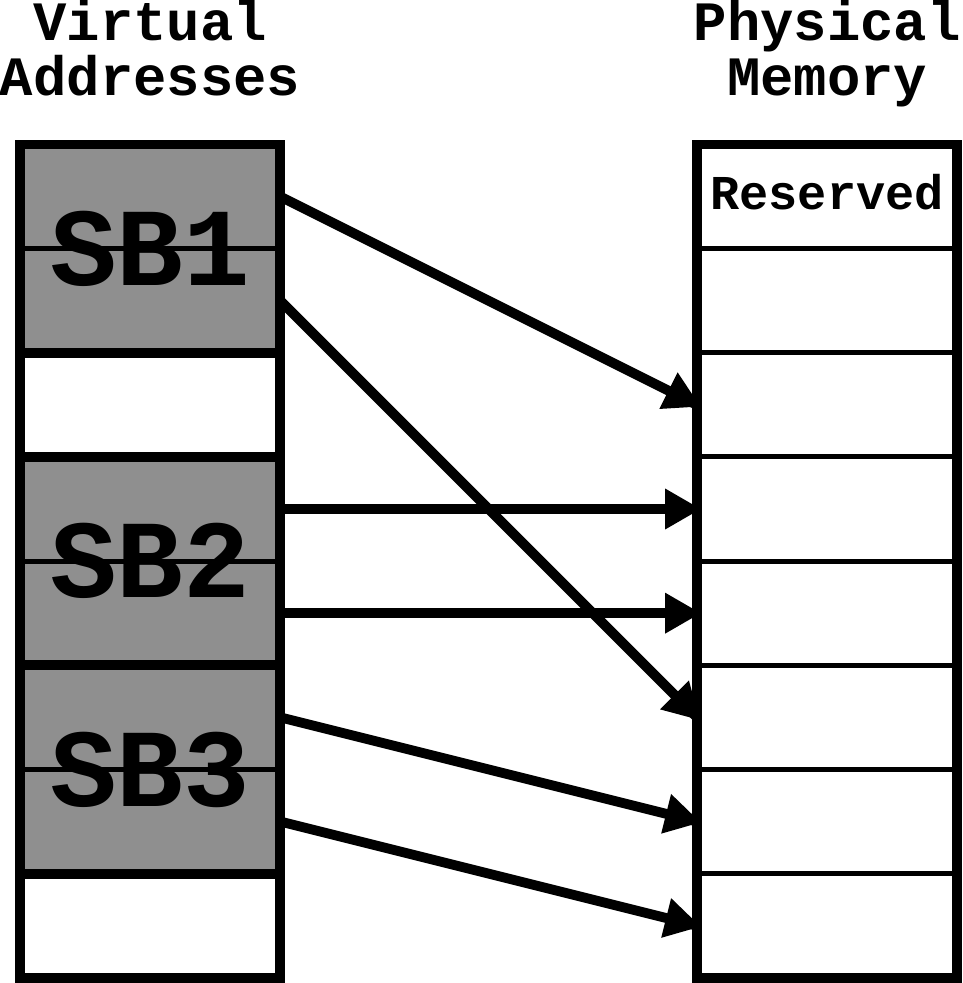}
		\caption{Before}
		\label{fig:remapa}
	\end{subfigure}
	\begin{subfigure}{0.49\linewidth}
		\includegraphics[width=0.95\columnwidth]{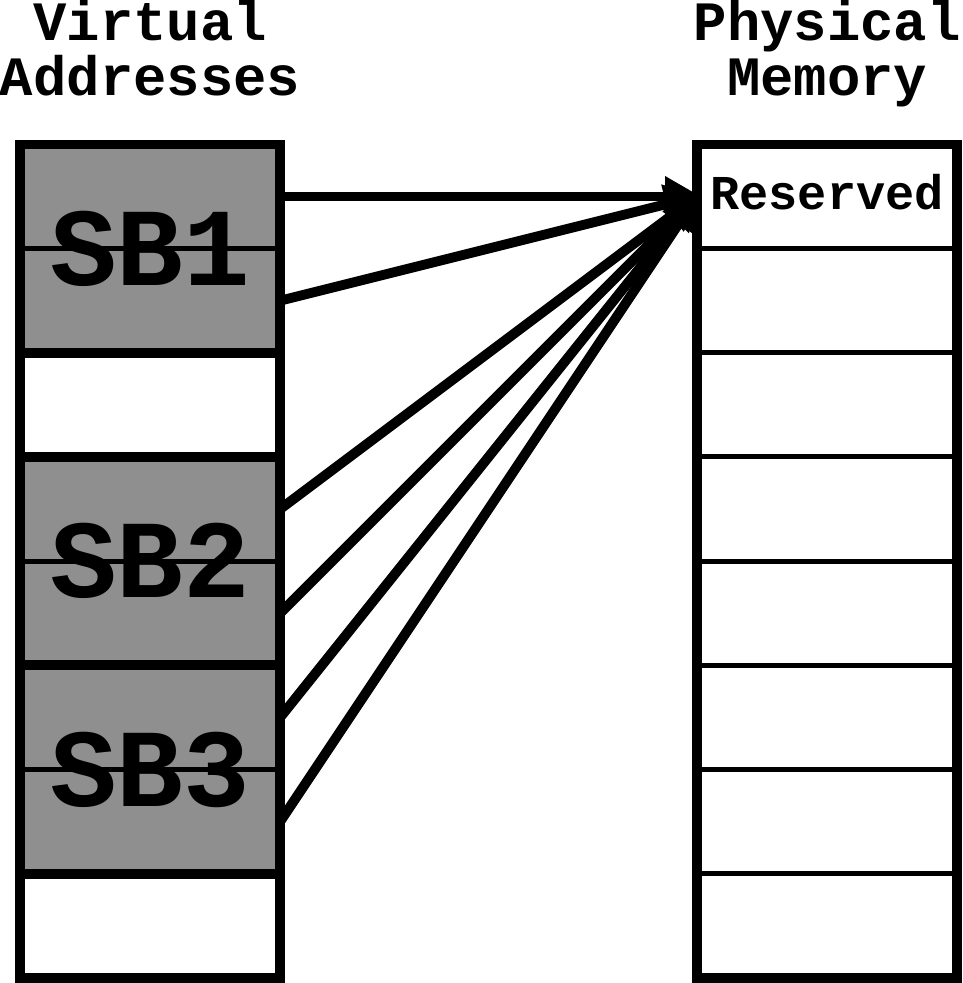}
		\caption{After}
		\label{fig:remapb}
	\end{subfigure}
        \caption{Memory mappings before and after the remapping process}
	\label{fig:remap}
\end{figure}

If we take a closer look to this problem, taking into account the
virtual memory system, we can observe that what actually needs to
remain accessible is the address range of the superblocks marked as
persistent and not the backing physical memory, as there is no
requirement regarding accessing the contents of the reclaimed
memory. Thus, the problem can be solved if we can release the physical
memory associated with such superblocks but maintain the addresses
range accessible.\footnote{Note that now we are considering again that
  all superblocks can become empty, i.e., ready to be released to the
  operating system.} To do so, we can remap the address range of a
persistent superblock becoming empty into a default pre-reserved
frame. Thus, independently of how many empty superblocks we have, they
will just consume a single frame of physical memory. This single frame
could even be a frame already in use by the process, as long as we can
ensure it will remain accessible throughout the lifetime of the
process. Figure~\ref{fig:remap} illustrates this remapping process. In
Fig.~\ref{fig:remapa} we show multiple persistent superblocks using 2
pages each, with each page mapped to a different frame, and in
Fig.~\ref{fig:remapb} we show how the superblocks can be remapped in
order to release all their frames while keeping the access to them
valid.

However, we need to be careful as the virtual address space is an
abundant but limited resource. So, some mechanism to recycle the
virtual addresses of the remapped superblocks still needs to be
used. But this is almost already done by LRMalloc when it needs to
recycle the \emph{descriptors} that contain the metadata of a
superblock. Remember that when a non-persistent superblock becomes
empty, the superblock is unmapped and the descriptor is added to the
recycling pool. Later, when a new superblock is requested, first, a
descriptor is obtained from the recycling pool, then a superblock is
mapped from the OS, and finally the metadata in the descriptor is
rewritten with the metadata of the new superblock. So, if we use
instead the address range stored in the descriptor obtained from the
recycling pool to map the new superblock, we are effectively recycling
the virtual address space by piggy backing on the descriptor. In the
actual implementation, we added an additional recycling pool with this
mechanism, which we give priority to obtain blocks from, and keep the
original for descriptors originated from non-persistent
superblocks. The reason for the second pool will become clearer in
section~\ref{sect:limitations}.

For the actual remapping process, we propose 2 methods. The first
method is to advise the operating system that the memory will not be
needed. In Linux, this is accomplished by the use of the
\emph{madvise()} system call with the \emph{MADV\_DONTNEED} flag,
which reverts the memory mapping to a state similar to when the
superblock was first allocated, i.e., all pages are mapped to a single
copy on write zero filled frame. This frees all physical memory
previously associated with the map until it is written again. Note
that reads to these ranges of memory do not cause a page fault, but
only an actual read from the zero filled frame. With this method, when
we get a descriptor from the recycling pool, we do not need to do any
extra work for remapping as the original address range is already
valid and ready to use.

This first method has the advantage of being simpler and more
efficient, but has two main disadvantages. One disadvantage is that
even though this system call and flag are defined in the POSIX
standard, the standard itself does not imposes the behavior observed
in Linux, which makes this method not portable. Another disadvantage
is that some optimistic access derived methods, like
VBR~\cite{sheffi21}, use DWCAS (Double-Width Compare-and-Swap) on
reclaimed memory, even though the DWCAS is certain to
fail\footnote{It uses tagged pointers as an ABA prevention
  mechanism.} as otherwise it would lead to corruption, the operating
system is unable to ascertain that and faults a frame in through the
copy on write mechanism. This does not cause a correctness issue but
could lead to some memory leaking, as some pages would be reserved for
unallocated superblocks.

The second method is to use the shared memory mechanism. We start by
defining a shared memory region and then, when we want to deallocate a
superblock, we map its address range to the shared memory region. We
can choose a size for the shared memory region that varies from the
size of a page to the size of a superblock, which can lead to
different performance trade-offs as we need one system call to do the
remap if we choose the size of a superblock, two system calls if we
choose half the size of a superblock, and so on. Note that the
physical memory associated with the shared memory region could be used
to store something useful in the meantime. For example, it could be
used to store the \emph{descriptors}. Later, to reuse the virtual
range of the superblock we need to remap it again to new memory. Note
that this remap only requires one system call, independently of the
size of the shared memory region. In Linux, this method is
accomplished with the use of the \emph{mmap()} system call with the
flags \emph{MAP\_FIXED} and \emph{MAP\_SHARED} to release the physical
memory, and \emph{MAP\_FIXED}, \emph{MAP\_PRIVATE} and
\emph{MAP\_ANON} to reuse the superblock.

Although this method might look a bit abusive, it is supported by the
POSIX standard. However, this support is not explicit and, in Linux,
the memory statistics go haywire, as it counts all the ranges mapped
to the shared mapping into the resident set size (RSS) of the process,
even though it only uses one shared mapping of physical memory. This
method can also be used in other operating systems outside the POSIX
world, and does not lead to memory leakage when CAS instructions are
used on reclaimed memory. It requires extra system calls but we were
not able to measure any performance degradation caused by them.


\section{Limitations}
\label{sect:limitations}

The LRMalloc memory allocator uses a size class allocation strategy,
which means that allocations up to a reasonable size (16KiB) are
handled through this mechanism. For all size class allocations,
LRMalloc uses superblocks of the same size (2MiB), which simplifies
our remapping logic as we can reuse retired supeblock addresses to
different size classes. This is ideal in most scenarios, as most
allocations fall into the size class range. However, for allocations
larger than the biggest size class, it requires a different
mechanism. For such allocations, LRMalloc relies directly on the
operating system, as other lock-free memory allocators
do~\cite{michael04alloc,gidenstam10,seo11,li20}. Relying on the
operating system for large allocations does not meaningfully impact
performance as this kind of allocations are uncommon. Large
allocations work similarly to size class allocations, but the thread
caches are skipped and a superblock with the exact size needed is
mapped to satisfy the allocation.

This way of dealing with large allocations is not ideal, as it
requires a different mechanism in order to recycle the range of
virtual addresses of such allocations. In this regard, we have chosen
to restrict the persistent memory allocation to sizes that are
compatible with the size classes. This is not a problem in most
situations as lock-free data structures tend to either use small
allocations for their internal structure, or the large allocations
last the lifetime of the data structure and as such need no
reclamation, one example being Michael's lock-free hash
tables~\cite{michael02lfht}. The exceptions are lock-free hash maps
that use large arrays that are resizable, as during the resizing
process they need to allocate a new array and reclaim the old one.
Data structures with these mechanisms are rather uncommon as the
resizing processes tend to be complex and synchronization heavy, which
leads to performance loss. As such, we leave the resolution of this
limitation to future work.

This limitation is also the reason why we need another recycling pool
for descriptors when a superblock becomes \emph{empty}. If the
superblock is not marked as persistent\footnote{Note that only
  superblocks used for size class allocations can become persistent.},
the superblock is unmapped and the descriptor is added to the pool
with the original behavior. If the superblock is marked as persistent,
we remap the superblock as shown in the previous section and add the
descriptor to the new pool. When we need a new descriptor we try to
obtain one using the following priority: (i) the new pool that already
has the virtual range of the superblock associated with it and as such
is only compatible with superblocks intended for size class
allocations; (ii) the original pool that has generic descriptors; and
finally, (iii) we allocate a new descriptor. We only go down the
priority list if either the pool is incompatible or is exhausted.


\section{Experimental Results}

In order to evaluate the impact of our changes to the OA method, we
compare the results of our two implementations of the OA method, the
one with warning-bits and the one with the monotonic global variable,
against the original OA method, and against no reclamation, in which
memory is never reclaimed, reused or freed. From this point onwards,
we will refer to our simplified OA method with the warning-bit per
thread as \emph{OA-BIT}, the alternative with the monotonic global
variable as \emph{OA-VER}, and the no reclamation alternative as
\emph{NR}.


\subsection{Methodology}

The hardware used was a machine with 2 AMD Opteron(TM) Processor 6274
with 16 cores each, 16KiB of L1 cache per core, 2MiB of L2 cache per
pair of cores and 12MiB of usable shared L3 cache per CPU. It has a
total of 32GiB of DDR3 memory.

We benchmarked the four methods with the commonly used Michael's
lock-free hash tables~\cite{michael02lfht} and Harris-Michael's
lock-free linked lists~\cite{michael04}. For all benchmarks, we use
LRMalloc as the memory allocator, and although for our simplified
versions it uses the new \emph{palloc()} procedure for allocation, for
both the original OA and no reclamation it uses the regular
\emph{malloc()} procedure. Note that the OA method only uses the
allocator to create its memory pool before the benchmark begins and
performs no allocations during the benchmark itself.

The benchmarks were run with varying ratios of searches, inserts and
removes, but we kept the ratio between inserts and removes at 1:1 in
order to keep the size of the data structure constant throughout the
benchmark. For linked lists, we ran the benchmarks with 5K nodes
pre-inserted. For hash tables, we used both 10K and 1M nodes and a
load factor of 0.75. The results are the mean of 10 runs of 1 second
each, and we show the results in the form of throughput (number of
operations per second) for every combination of threads from 1 to 32.

For all these experiments, we are not showing comparisons between the
different approaches to memory remapping because we were unable to
measure any difference in performance (outside a margin of error)
between keeping the memory in the allocator, advising the operating
system with \emph{MADV\_DONTNEED} and remapping with a shared memory
region.


\subsection{Results}

\begin{figure*}
	\begin{subfigure}{0.49\linewidth}
		\includegraphics[width=\textwidth]{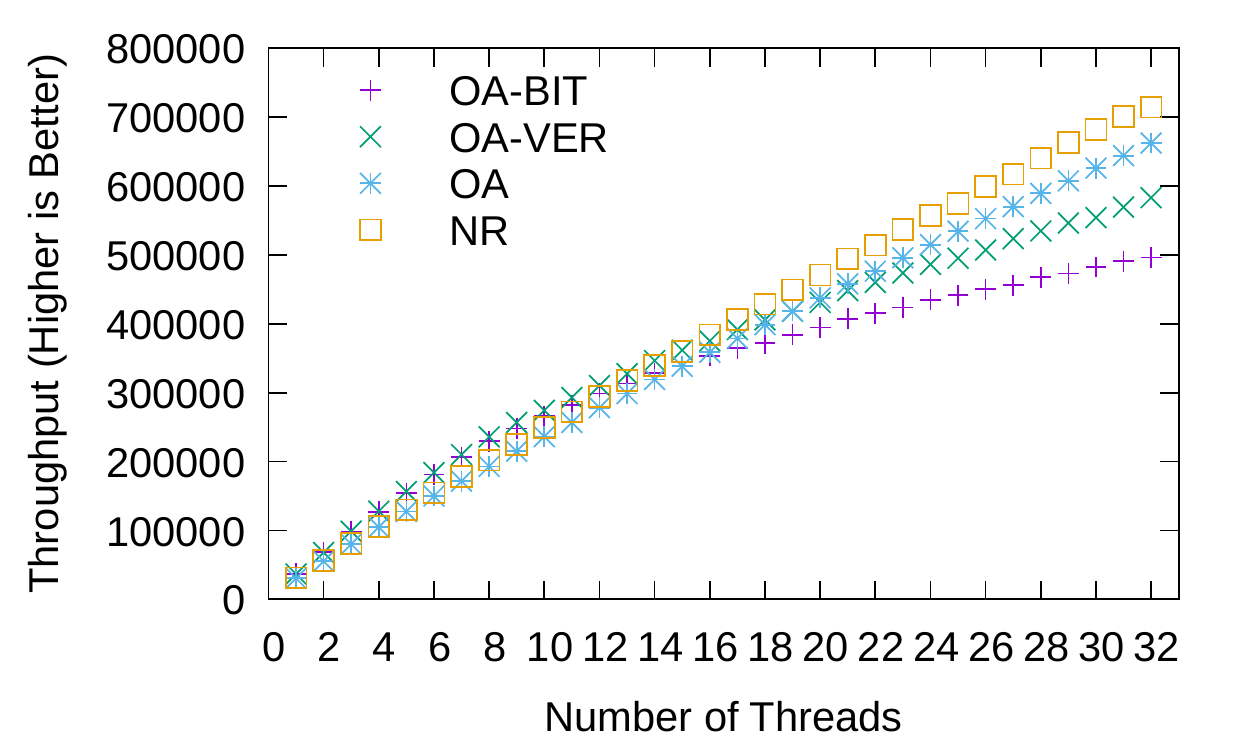}
		\caption{50\% inserts and 50\% removes}
		\label{fig:l5k0}
	\end{subfigure}
	\begin{subfigure}{0.49\linewidth}
		\includegraphics[width=\textwidth]{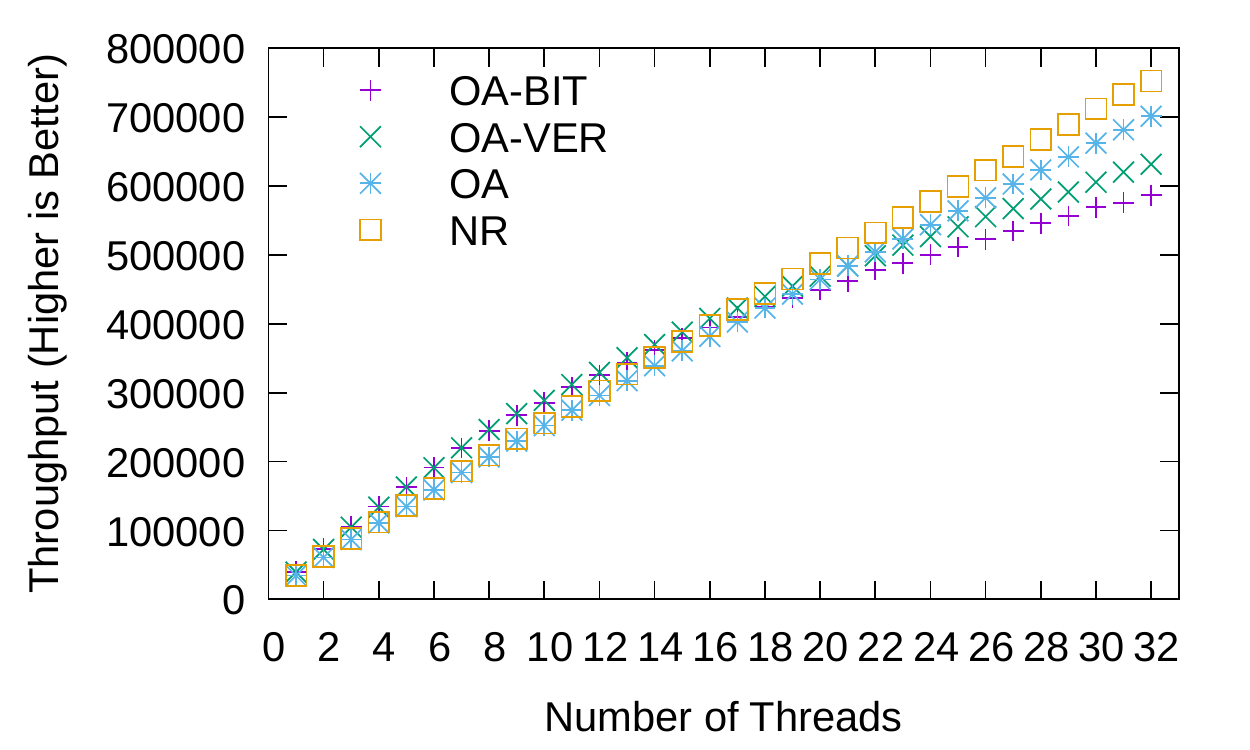}
		\caption{50\% searches, 25\% inserts and 25\% removes}
		\label{fig:l5k5}
	\end{subfigure}
	\caption{Linked lists with 5K nodes}
	\label{fig:l5k}
\end{figure*}

\begin{figure*}
	\begin{subfigure}{0.49\linewidth}
		\includegraphics[width=\textwidth]{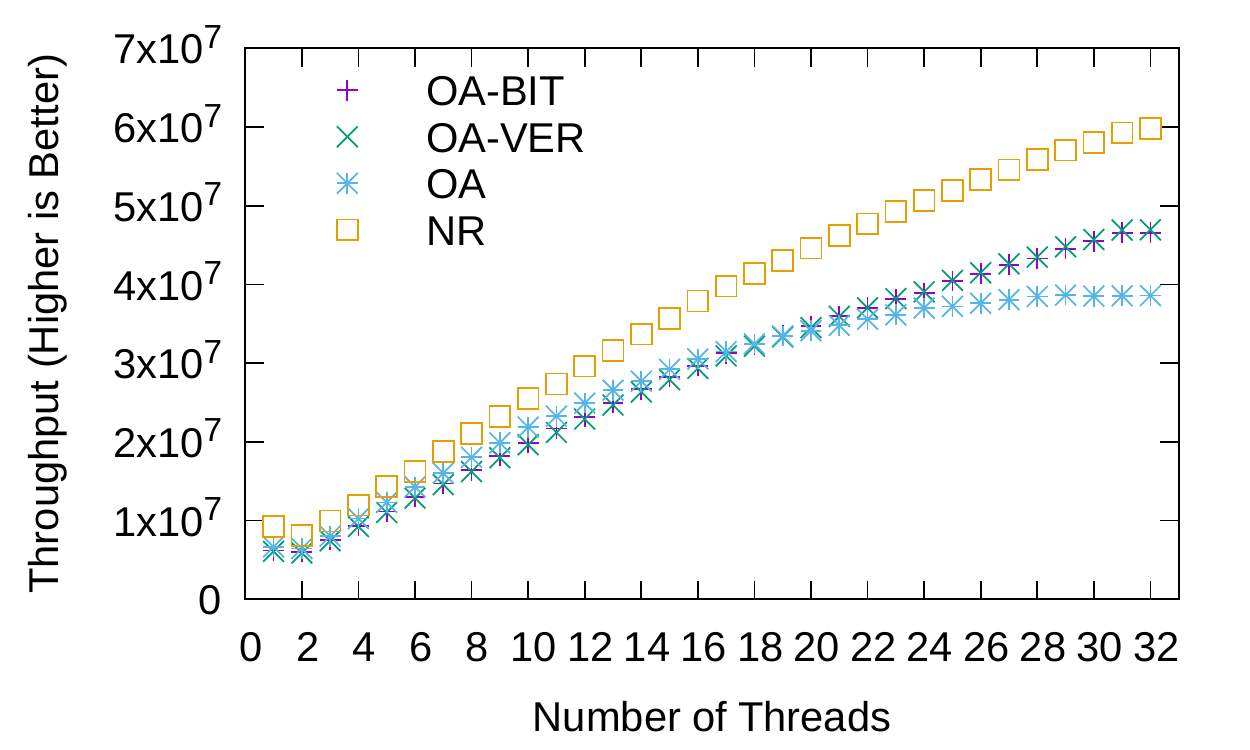}
		\caption{50\% Inserts and 50\% Removes}
		\label{fig:h10k0}
	\end{subfigure}
	\begin{subfigure}{0.49\linewidth}
		\includegraphics[width=\textwidth]{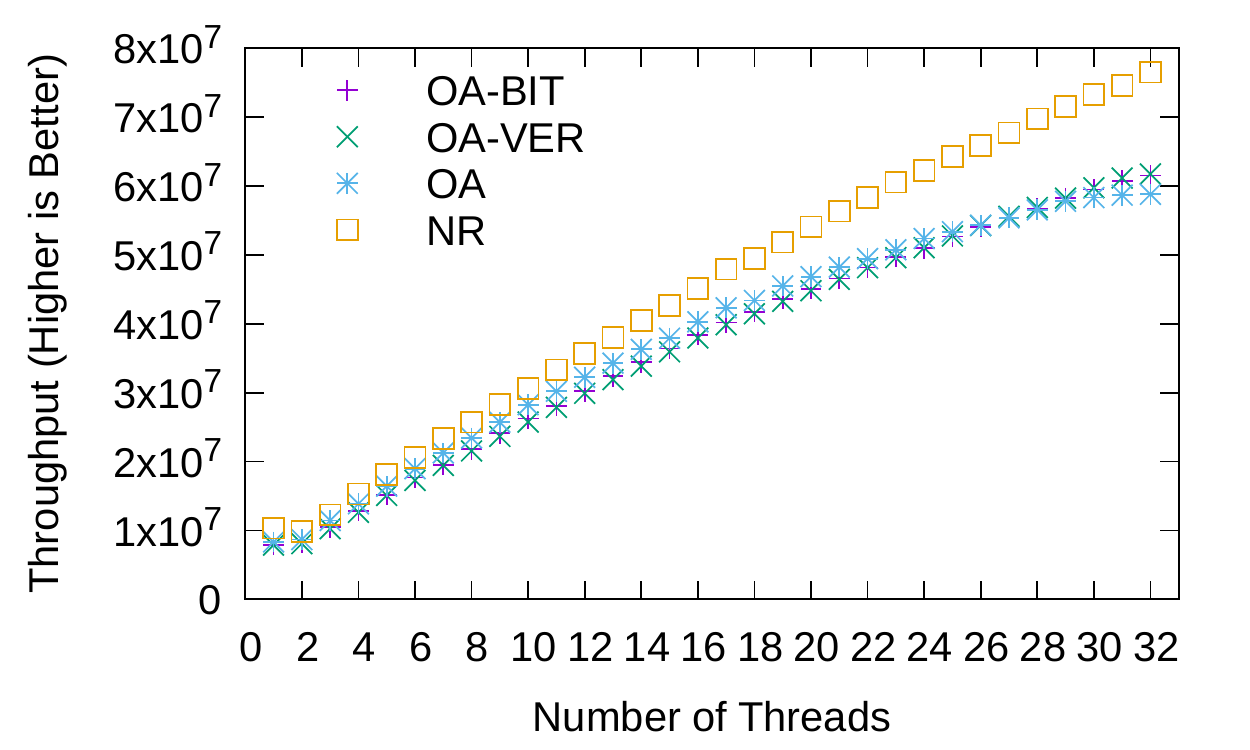}
		\caption{50\% Searches, 25\% Inserts and 25\% Removes}
		\label{fig:h10k5}
	\end{subfigure}
	\caption{Hash Table with 10K nodes}
	\label{fig:h10k}
\end{figure*}

\begin{figure*}
	\begin{subfigure}{0.49\linewidth}
		\includegraphics[width=\textwidth]{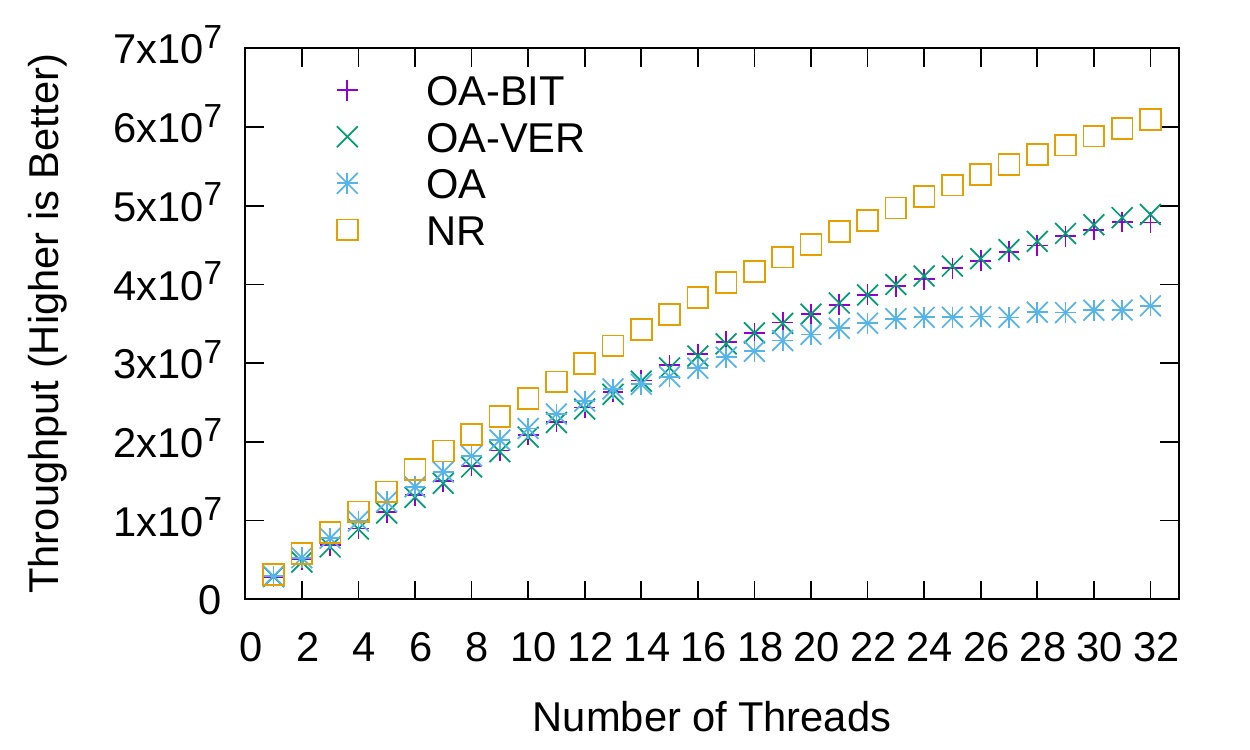}
		\caption{50\% Inserts and 50\% Removes}
		\label{fig:h1m0}
	\end{subfigure}
	\begin{subfigure}{0.49\linewidth}
		\includegraphics[width=\textwidth]{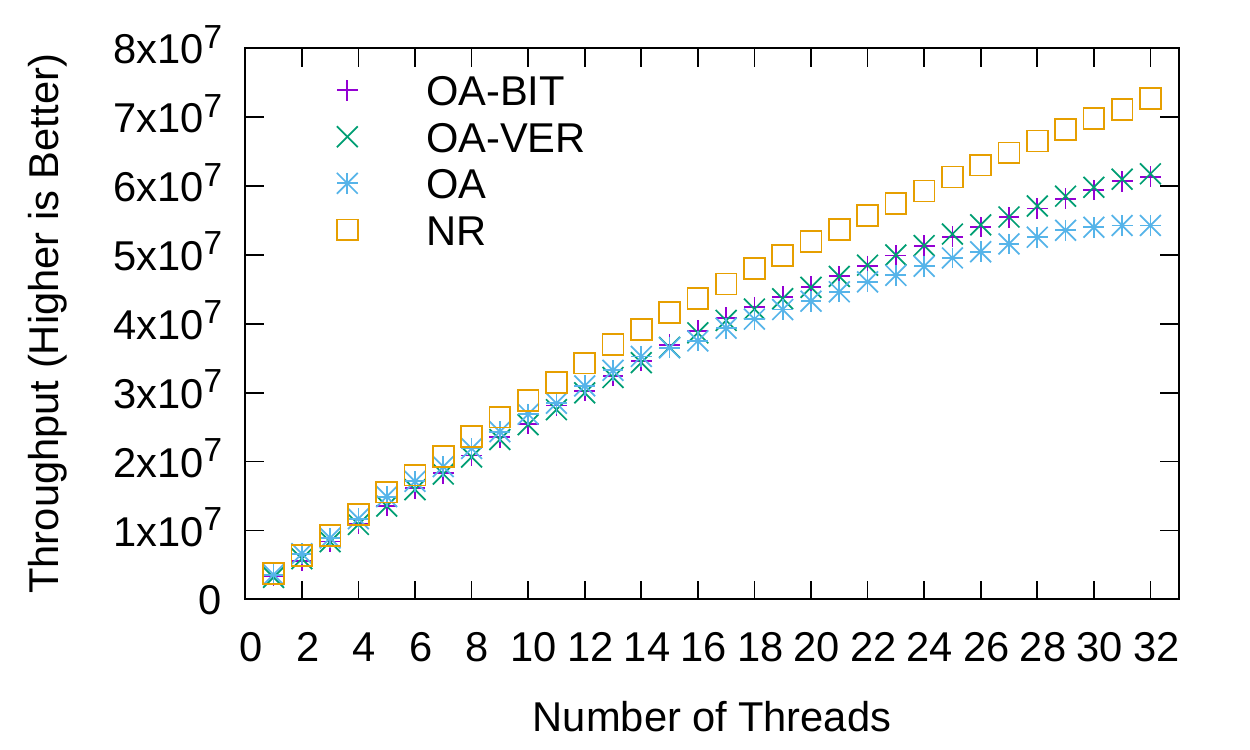}
		\caption{50\% Searches, 25\% Inserts and 25\% Removes}
		\label{fig:h1m5}
	\end{subfigure}
	\caption{Hash Table with 1M nodes}
	\label{fig:h1m}
\end{figure*}

Figure~\ref{fig:l5k} shows the results for the benchmark using linked
lists with 5K nodes pre-inserted. Figure~\ref{fig:l5k0} shows the case
with only modifying operations (50\% inserts and 50\% removes) and
Fig.~\ref{fig:l5k5} shows a more balanced set of operations (50\%
searches, 25\% inserts and 25\% removes). Figures~\ref{fig:h10k}
and~\ref{fig:h1m} then show the results for the benchmarks using hash
tables with 10K nodes and 1M nodes, respectively. For both benchmarks,
we also have the case with only modifying operations (50\% inserts and
50\% removes) and with a more balanced set of operations (50\%
searches, 25\% inserts and 25\% removes).

For linked lists with only modifying operations, the OA-VER method
shows significant improvements to the OA-BIT method due to its ability
to fire less warnings. This effect is somewhat reduced for linked
lists with 50\% searches, as there are less removes, and becomes
negligible in both benchmarks using hash tables (Figs.~\ref{fig:h10k}
and~\ref{fig:h1m}) due to the much shorter chains.

For low amounts of threads, we can see that both OA-BIT and OA-VER
outperform the OA and even the NR method for linked lists. This
happens because with low amounts of threads our methods use less
memory, keeping most of the memory used in lower level caches. With
increasing number of threads, our two methods start using more memory
due to the per thread caches of LRMalloc and thus loose this advantage
to the OA method that has a memory pool of a fixed size and to the NR
method that suffers from less overhead caused by synchronisation
between the many threads. A memory allocator with different
characteristics could show a different behavior here. Linked lists are
an unresting example to study the behavior of the system but they are
not the ideal tool when performance matters due to their asymptotic
complexity characteristics.

The benchmarks using hash tables show a kind of inversion of the
results. In general, the OA method shows slightly better performance
than our methods for low amounts of threads, but a clear lack of
scalability for higher thread counts. Here, since we are working with
much higher throughputs and larger amounts of memory, the weight of
synchronization becomes much more relevant compared to memory usage
and thus cache locality. The fixed size of the memory pool in the OA
method proves detrimental as it requires much more recycling phases as
the throughput and thread counts increase, causing synchronization to
increase as well. In both our methods, we do not suffer from these
drawbacks as the thread caches in the allocator and private limbo
lists allow for less synchronization and thus better scalability.

Please remember that the main contribution of this paper is the added
ability of releasing memory to the memory allocator/operating system
and the simplification of the memory reclamation method, not the
performance and scalability gains, even thought they are welcome.


\section{Related and Future Work}

Since the proposal of the OA method, some other proposals have been
developed focusing on making OA easier to use and compatible with more
data structures. One such example is the \emph{Automatic Optimistic
  Access} (AOA) method~\cite{cohen15aoa}, which allows the data
structure programmer to forego the retire call by making use of
garbage collector like techniques. A second example is the \emph{Free
  Access} (FA) method~\cite{cohen18} that requires the programmer to
annotate the data structure functions, which then, through a
combination of garbage collection techniques and compiler steps, is
able to apply OA like memory reclamation to the data structure without
the need for it be written in a normalized
form~\cite{timnat14}. Another example is the VBR
method~\cite{sheffi21} that is able to extend OA to write operations
through the use of DWCAS (Double-Width Compare-and-Swap) with tagged
pointers.

We already discussed how our modifications to LRMalloc can be
compatible with the optimistic DWCAS of VBR, so we leave it to future
work the simplification and adaptation of VBR in order to also make it
able to release memory back to the memory allocator/operating
system. We could also use the extended LRMalloc in order to allow a
dynamic resizing of the memory pool (in a garbage collector like
manner) both in the AOA and FA methods, allowing the memory pool to be
shrunk by releasing it to the memory allocator/operating system. Our
results for the linked list benchmark show that this could also lead
to performance improvements.

Further work also includes the removal of the limitation discussed in
Section~\ref{sect:limitations}, which requires a mechanism capable of
splitting and coalescing virtual address ranges in a lock-free manner.


\section{Conclusion}

Starting from a lock-free general purpose memory allocator named
LRMalloc, we showed how to extend it to support the memory model
required by the OA memory reclamation method in such a way that we can
guarantee memory allocations to be readable even after we free such
allocations. We were able to eliminate the major drawback of the OA
method while ensuring that it remains one of the most efficient memory
reclamation methods. While doing so, we were also able to simplify the
implementation of the OA method, and obtain results showing
performance improvements.


\bibliographystyle{acm}
\bibliography{references}
\end{document}